\documentclass{elsarticle}
\pdfoutput=1

\usepackage{amsmath}
\usepackage{amssymb}
\usepackage{soul}
\usepackage{fullpage}
\usepackage{graphics}
\usepackage{graphicx}
\usepackage{xspace}
\usepackage{hyperref}
\usepackage{textcomp}
\usepackage{xfrac}
\usepackage{color,xcolor}

\newcommand{\harm}{\texttt{Harm3D}\xspace}
\newcommand{\IGM}{\texttt{IllinoisGRMHD}\xspace}
\newcommand{\ET}{\texttt{Einstein Toolkit}\xspace}
\newcommand{\mumps}{\texttt{MUMPS}\xspace}
\newcommand{\eigen}{\texttt{eigen}\xspace}
\newcommand{\NH}{\texttt{NewHorizons}\xspace}
\newcommand{\ud}{\mathrm{d}}
\newcommand{\tdyn}{t_{\rm dyn}}
\newcommand{\tinysub}{\scriptscriptstyle}
\newcommand{\onehalf}{\sfrac{1}{2}}
\newcommand{\nghost}{n_{\mathrm{ghost}}}

\begin{document}

\begin{frontmatter}

\title{Numerical generation of vector potentials from specified magnetic fields}

\author[RIT-Z]{Zachary J. Silberman\corref{corrauth}}
\cortext[corrauth]{Corresponding author}
\ead{zjs2405@rit.edu}

\author[WVU-TI]{Thomas R. Adams}
\ead{adams.tom.r@gmail.com}

\author[RIT-J]{Joshua A. Faber}
\ead{jafsma@rit.edu}

\author[WVU-TI,WVU-Z]{Zachariah B. Etienne}
\ead{zachetie@gmail.com}

\author[WVU-TI]{Ian Ruchlin}
\ead{ianruchlin@gmail.com}

\address[RIT-Z]{Center for Computational Relativity and Gravitation, School of Mathematical Sciences, and School of Physics and Astronomy, Rochester Institute of Technology, Rochester, New York 14623}
\address[WVU-TI]{Department of Mathematics, West Virginia University, Morgantown, West Virginia 26506}
\address[RIT-J]{Center for Computational Relativity and Gravitation, School of Mathematical Sciences, Rochester Institute of Technology, Rochester, New York 14623}
\address[WVU-Z]{Center for Gravitational Waves and Cosmology, West Virginia University, Chestnut Ridge Research Building, Morgantown, West Virginia 26505}

\begin{abstract}
Many codes have been developed to study highly
relativistic, magnetized flows around and inside compact
objects. Depending on the adopted formalism, some of these codes
evolve the vector potential $\mathbf{A}$, and others evolve the
magnetic field $\mathbf{B}=\nabla\times\mathbf{A}$ directly. Given that
these codes possess unique strengths, sometimes it is desirable to
start a simulation using a code that evolves $\mathbf{B}$ and complete
it using a code that evolves $\mathbf{A}$. Thus transferring the data
from one code to another would require an inverse curl
algorithm. This paper describes two new inverse curl techniques in
the context of Cartesian numerical grids: a cell-by-cell method,
which scales approximately linearly with the numerical grid, and a
global linear algebra approach, which has worse scaling properties
but is generally more robust (e.g., in the context of a magnetic
field possessing some nonzero divergence). We demonstrate these
algorithms successfully generate smooth vector potential
configurations in challenging special and general
relativistic contexts.
\end{abstract}


\end{frontmatter}

\section{Introduction}
\label{intro}

One of the major concerns in numerical evolutions of magnetic fields,
often as a component of a broader magnetohydrodynamic (MHD)
simulation, is ensuring that the $\nabla\cdot\mathbf{B}=0$ constraint of
Maxwell's equations remains satisfied.  If an MHD code cannot maintain
this condition, the resulting output may quickly become unphysical due
to the introduction of monopoles.

Different numerical algorithms have been developed to ensure that
the ${\nabla\cdot\mathbf{B}=0}$ constraint is maintained throughout the
simulation domain, both in space and in time.  For grid-based Eulerian
codes---the application of interest for the authors---these techniques
include constrained transport schemes~\cite{Balsara:1999aa,Toth:2000},
in which the electromagnetic induction equations are carefully
rewritten to ensure that if the divergence-free condition is satisfied
initially, it is automatically satisfied at all times. An
equivalent technique in the case of uniform-resolution, single-patch
grids involves evolving the magnetic vector potential $\mathbf{A}$ as a
fundamental variable rather than the magnetic field $\mathbf{B}$ (see, e.g.,
\cite{Etienne:2015}).  In the latter case, the fields can be recovered
from the potential, which is defined such that
$\mathbf{B}=\nabla\times\mathbf{A}$.  The divergence of a curl is zero, so
the resulting magnetic field computed from the vector potential at any
point in space and time is \textit{automatically} divergence-free.

Often we would prefer to evolve the vector potential $\mathbf{A}$, as
\textit{any} interpolation strategy can be applied to the
$\mathbf{A}$-fields without introducing violation to the
${\nabla\cdot\mathbf{B}=0}$ constraint beyond roundoff-level. To this end,
with an eye towards scenarios that arise in numerical relativity, we
have in mind MHD evolutions on Cartesian grids using the
\IGM~\cite{Etienne:2015} code within the \ET. \IGM is geared toward
simulations of general relativistic MHD (GRMHD) fluid flows in
highly-dynamical spacetimes, leading potentially to electromagnetic
counterparts to gravitational wave signals observable by Advanced
LIGO.

However, \IGM requires that $\mathbf{A}$ fields be specified at all grid
points, and in some astrophysically-relevant contexts we are given
only magnetic field data, with $\mathbf{A}$ unspecified.  Thus, we need a
way to generate an $\mathbf{A}$ field corresponding to a specified
$\mathbf{B}$ field.  This is a key stumbling block, in particular for
contexts in which \IGM is the \textit{only} open-source code available
to continue GRMHD evolutions started by other, $\mathbf{B}$-field based
codes like \harm \cite{Gammie:2003rj}, which may be used, for
instance, to perform long-term MHD simulations for spacetimes that
either evolve slowly or retain a high degree of symmetry.

Unfortunately, generating a vector potential configuration
corresponding to a given magnetic field can be challenging, especially
if the magnetic field configuration can, in practical terms, only be
represented numerically. To do so, an ``inverse-curl'' operator would
be necessary, potentially including the specification of an
electromagnetic gauge to uniquely define the resulting solution.

Several analytic and semi-analytic methods for calculating the inverse
curl have been derived, but none are particularly convenient for
computing the vector potential of a grid of points using Cartesian
coordinates.  Sahoo \cite{Sahoo:2008}, for instance, derives
fully-analytic expressions for inverse vector operations, including
the inverse curl.  This expression involves integrals of the
components of $\mathbf{B}$ with respect to the coordinates and thus
assumes a simple analytic form of $\mathbf{B}$ is known, which would
greatly reduce the computational cost of this integration. In
numerical work, we do not typically have this luxury.  Webb et
al. \cite{Webb:2010} also find an integral expression for $\mathbf{A}$ in
terms of $\mathbf{B}$.  Their formula, in addition to requiring a smooth
magnetic field, involves integrals along line segments from an origin
to the point of interest, which is not computationally feasible for a
simulated grid of points.  In addition, the vector potential derived
numerically in this method has the unfavorable quality that it is
path-dependent. 

Other methods of calculating the inverse curl have been found for
non-Cartesian coordinate systems.  Edler \cite{Edler:2010} derives
expressions for the spherical harmonic coefficients of the magnetic
fields. From the coefficients, the vector potential can be directly
computed.  However, we do not know of basis functions for Cartesian grids that are as useful 
for this purpose as spherical harmonics are for spherical grids.  Yang et
al.~\cite{Yang:2013} apply a linear-algebra-based technique to extract
the 12 integrals of the vector potential along the edges of a volume
from the six values of $\mathbf{B}$ on the boundaries of that volume.
The values of $\mathbf{A}$ can then be derived from these integrals using
sinusoidal functions.  The downside to this method is that it assumes
a single volume, not small cells within a volume, so there is no way
to ensure consistency from one cell to another.

Several groups have adopted finite element analysis to solve practical
problems involving magnetic fields and vector potentials, and in the
process helped to develop the technique further.  For example,
Demerdash, Nehl, and Fouad \cite{Demerdash:1980} use a tetrahedral
element as their basic element.  The magnetic field values are defined
at the centers of the elements, and they are constant within each
tetrahedron.  Each element has four values for the vector potential,
at the four vertices.  Biro, Preis, and Richter \cite{Biro:1996} find
that using edge-based finite elements for the magnetic vector
potential is more accurate and numerically stable than using nodal
elements.  In addition, they conclude that using the Coulomb gauge
for the vector potential makes the convergence of the solution for
edge-based elements much worse.  
Even so, the methods of finite element analysis are still relevant to
our work: the ability to split a domain into smaller elements, of
various possible shapes and sizes, is crucial to calculating the
vector potential in the interior of a region, not just on its boundary. 

In this paper, we describe two techniques \cite{codes:web} that can be 
used to perform the inverse curl operation on a staggered numerical grid,
including a direct, cell-by-cell approach, as well as
a global method involving large-scale sparse linear algebra solvers.
The results may be compared using a variety of measures to describe
the smoothness of the resulting solution, and we can compare their
performance when used to generate initial data for numerical runs.
Our paper is organized as follows: In Sec. \ref{stagger}, we describe
the staggered grids adopted to solve the problem.  Sec. \ref{solutions}
outlines the numerical algorithms we have developed to calculate the
inverse curl.  Performance and scaling of our codes are reviewed in
Sec. \ref{performance} and the validation tests are presented in
Sec. \ref{tests}.  Sec. \ref{conclusion} concludes the paper.

\section{Vector potentials and magnetic fields on staggered grids}
\label{stagger}

\subsection{Geometry of staggered grids}
\label{staggergeo}

Our grids adopt a fixed step size in each of the three Cartesian
directions, and ignore complications that arise in relativity from
non-constant spatial metrics.  In practice, the magnetic constraint
equations can always be written in terms of flat-space
divergences and curls of quantities, so this assumption results in no
loss of generality.  

\begin{figure}[h!]
\centering
\includegraphics[width=4.0in]{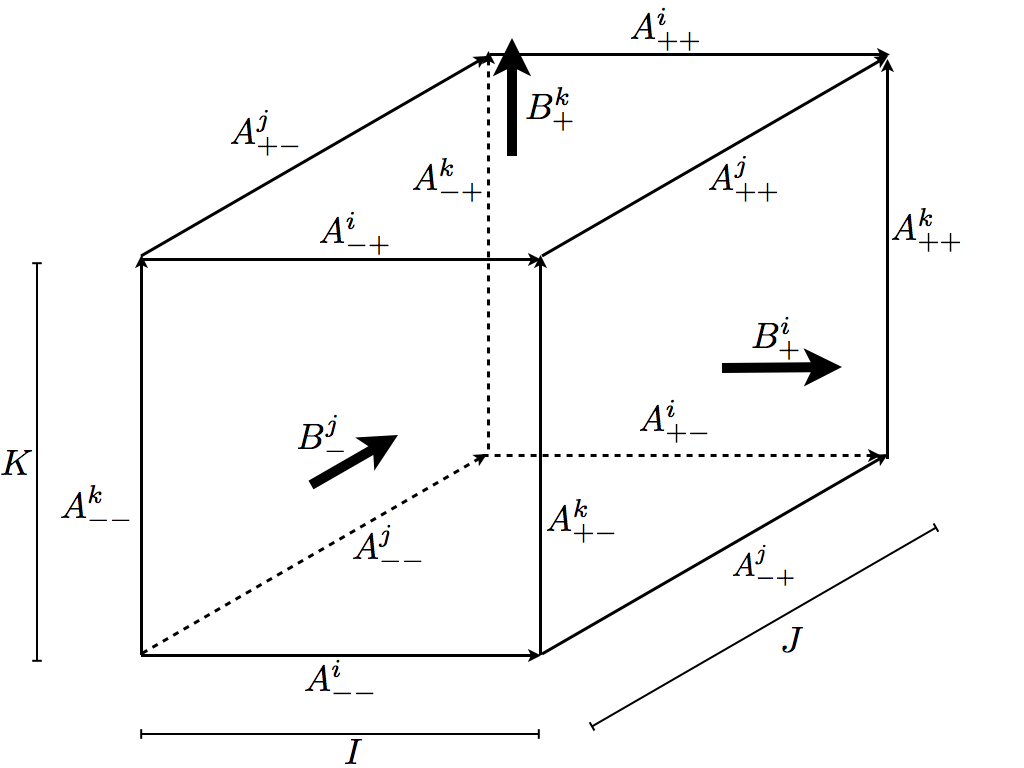}
\caption{Example numerical grid cell, showing the locations at which
  various quantities are defined.  Hydrodynamic 
  quantities are defined at cell centers, magnetic fields at the
  centers of the cell faces (for clarity, only three of the six values
  for this cell are labeled), and magnetic vector potentials at the
  centers of cell edges.  Due to the symmetry of
  the problem, $(i,j,k)$ may be taken as any positively oriented cycle
  of the elements $(x,y,z)$ See Eqs.~\eqref{eq:Bloc},
  \eqref{eq:Aloc}, and the surrounding text for an explanation of the
  index conventions for the fields.}
\label{fig:exgridcell}
\end{figure}

Following the standard approaches used by both
constrained transport and vector potential evolution codes, we
assume staggered grids for various quantities.
Specifically, all hydrodynamic quantities, including fluid pressures
and velocities, are known at points represented in
Fig. \ref{fig:exgridcell} as the centers of grid cells.  We note that
we reverse the standard conventions about grid-cell locations
for visual clarity; one would typically describe integer-indexed
quantities as the \emph{vertices} of the grid, rather than the
\emph{centers}.  Similarly, it reverses the notions of grid faces,
whose values are represented in our presentation as those with two
integer indices and one half-integer index, and grid edges, which here
have two half-integer indices and one integer index.  In order to
convert back to the standard picture, in which grid cells are shifted
by half a cell-width in each direction, one must interchange centers
with vertices and edges with faces.  Modulo this conversion, the
procedure below remains unchanged. 

For a numerical grid cell with index $(i,j,k)$ (where $i$ corresponds
to a unique Cartesian point $x$, $j$ a unique Cartesian point $y$, and
$k$ a unique Cartesian point $z$), the hydrodynamic
variable storage locations are defined as follows
\begin{equation}
\rho,P,\mathbf{v}\,:\,(i,j,k) \, .
\label{eq:hydroloc}
\end{equation}
In order to maintain divergence-free magnetic fields, we evaluate the
expression $\nabla\cdot \mathbf{B}$ at grid cell centers by shifting the
evaluation points for the magnetic field to cell faces,
\begin{equation}
B_{\tinysub{\pm}}^x:\left(i\pm\onehalf,j,k\right); \quad B_{\tinysub{\pm}}^y:\left(i,j\pm\onehalf,k\right); \quad B_{\tinysub{\pm}}^z:\left(i,j,k\pm\onehalf\right) \, .
\label{eq:Bloc}
\end{equation}
Then $\nabla\cdot\mathbf{B}$ at point $(i,j,k)$ is given by:
\begin{equation}
\nabla\cdot\mathbf{B} = \frac{B^x_{\tinysub{+}} - B^x_{\tinysub{-}}}{I} + \frac{B^y_{\tinysub{+}} - B^y_{\tinysub{-}}}{J} + \frac{B^z_{\tinysub{+}} - B^z_{\tinysub{-}}}{K} \, ,
\label{eq:divb0_ijk}
\end{equation}
where $I$, $J$, and $K$ represent the Cartesian $x$-, $y$-, and
$z$-directions grid spacings of a cell, respectively.  Performing the
calculation in this way, we obtain a second-order, centered
finite-differencing scheme. 

The divergence-free condition will be satisfied to roundoff error
\emph{automatically}, provided we define the vector potential
$\mathbf{A}$ at the edges of each cell with staggering given by
\begin{equation}
A^x_{\tinysub{\pm\pm}} : \left(i,j\pm\onehalf,k\pm\onehalf\right); ~ 
A^y_{\tinysub{\pm\pm}} : \left(i\pm\onehalf,j,k\pm\onehalf\right); ~ 
A^z_{\tinysub{\pm\pm}} : \left(i\pm\onehalf,j\pm\onehalf,k\right) \, ,
\label{eq:Aloc}
\end{equation}
where we note for reasons of cyclic symmetry that one should read the
$\pm$ subscripts for $A^y$ terms as representing the $z$-direction
offset and then the $x$-direction offset, rather than the other way
around.  Then the discretized formula for the curl is
\begin{equation}
B^i_{\tinysub{\pm}} = \frac{A^k_{\tinysub{\pm +}} - A^k_{\tinysub{\pm -}}}{J} - \frac{A^j_{\tinysub{+\pm}} - A^j_{\tinysub{-\pm}}}{K} \, ,
\label{eq:disccurl}
\end{equation}
where $(i,j,k)$ represent a positively oriented cycle of the elements
$(x,y,z)$---i.e., $(x,y,z)$, $(y,z,x)$, or $(z,x,y)$. 

Cancellation of the numerical divergence of $\mathbf{B}$ is guaranteed,
as the value of the vector potential at each edge is both added and
subtracted when evaluating Eq.~\eqref{eq:divb0_ijk}.  
For example, the curl condition on the top face of the cell
yields the expression
\begin{align}
B^z_{\tinysub{+}}  =& \;\frac{A^y_{\tinysub{++}} - A^y_{\tinysub{+-}}}{I} - \frac{A^x_{\tinysub{++}} - A^x_{\tinysub{-+}}}{J} \nonumber \\ 
B^z(i,j,\onehalf) =& \;\frac{A^y(i+\onehalf,j,k+\onehalf) - A^y(i-\onehalf,j,k+\onehalf)}{I} \nonumber \\
                &- \frac{A^x(i,j+\onehalf,k+\onehalf) - A^x(i,j-\onehalf,k+\onehalf)}{J} \, .
\label{eq:topcurl}
\end{align}

We note that for the purposes of bookkeeping, the quantities defined
above potentially have different dimensions because of the staggering.
If the grid of cells has dimension $L\times M\times N$ in the $x$-,
$y$-, and $z$-directions, respectively, then the numerical grid sizes of
the magnetic quantities are as follows:
\begin{align}
A^x &: L\times (M+1)\times (N+1) \, , &\quad B^x &: (L+1)\times M\times N \, , \nonumber \\
A^y &: (L+1)\times M\times (N+1) \, , &\quad B^y &: L\times (M+1)\times N \, , \nonumber \\
A^z &: (L+1)\times (M+1)\times N \, , &\quad B^z &: L\times M\times (N+1) \, .
\label{eq:ABsizes}
\end{align}

\subsection{Uniqueness and gauge choice}
\label{unique}

Determining a vector potential $\mathbf{A}$ that corresponds to a given
magnetic field $\mathbf{B}$ is an under-constrained problem, and
solutions are never unique. This is known as the gauge freedom of
$\mathbf{A}$. In general, if $\nabla\times\mathbf{A}=\mathbf{B}$, it is also
true that
\begin{equation}
\nabla\times (\mathbf{A}+\nabla\phi) = \mathbf{B} \, ,
\label{eq:Agradphi}
\end{equation}
for any scalar field $\phi$.  For numerical reasons, the methods
described here produce solutions in the Coulomb gauge, for which
$\nabla\cdot\mathbf{A}=0$; i.e., the vector potential is also
divergence-free.  Numerically, this is equivalent to the condition
\begin{align}
0 =& \left[\nabla\cdot\mathbf{A}\right](i+\onehalf,j+\onehalf,k+\onehalf) \nonumber \\
  =& \frac{A^x(i+1,j+\onehalf,k+\onehalf) - A^x(i,j+\onehalf,k+\onehalf)}{I} \nonumber \\
   &+ \frac{A^y(i+\onehalf,j+1,k+\onehalf) - A^y(i+\onehalf,j,k+\onehalf)}{J} \nonumber \\
   &+ \frac{A^z(i+\onehalf,j+\onehalf,k+1) - A^z(i+\onehalf,j+\onehalf,k)}{K} \, ,
\label{eq:Coulomb}
\end{align}
where the natural location to evaluate this expression is at the grid
vertices.

The Coulomb gauge is uniquely defined by requiring
regularity when considering infinite domains.
For finite domains, however, the Coulomb gauge solution is not unique,
as one may add in gradients of the solutions of the Laplace
equation that take the form $\mathbf{a}=\nabla(r^lY_{lm})$, where $r$ is
the spherical radius, $l$ is a non-negative integer, and $Y_{lm}$ is a
spherical harmonic function, without affecting either the curl or the
divergence of $\mathbf{A}$:
\begin{align}
\nabla\times\mathbf{a} = \mathbf{0} &\Rightarrow \nabla\times (\mathbf{A}+\mathbf{a})   = \nabla\times\mathbf{A} = \mathbf{B} \, , \nonumber \\
\nabla\cdot\mathbf{a}  = 0       &\Rightarrow \nabla\cdot(\mathbf{A}+\mathbf{a}) = \nabla\cdot\mathbf{A} = 0 \, .
\label{eq:curldiva}
\end{align}

Given that our domains are finite, we must therefore specify
additional boundary conditions for our numerical solutions. These
additional conditions are discussed below.

\section{Solution techniques}
\label{solutions}

In this section we describe two solution techniques to calculate the
vector potential.  We note that neither of these techniques are based
on the Helmholtz decomposition, in which the vector potential can be
calculated via:
\begin{equation}
\mathbf{A}_{\rm H}(\mathbf{r}) = \frac{1}{4\pi}\int_{\mathcal{V}} \frac{\nabla'\times\mathbf{B}(\mathbf{r}')}{|\mathbf{r}-\mathbf{r}'|}\ud V'
- \frac{1}{4\pi}\oint_{\partial \mathcal{V}} \frac{\hat{\mathbf{n}}'\times\mathbf{B}(\mathbf{r}')}{|\mathbf{r}-\mathbf{r}'|} \ud S' \, .
\label{eq:Ah}
\end{equation}
For magnetic fields that extend through the domain boundaries, the
second term in this equation is a non-zero surface term that is 
\emph{more difficult} to compute globally than either of the techniques 
we describe below.  In addition, a corresponding expression for a scalar 
potential $\Phi$ includes an additional surface term that must be 
eliminated via a gauge transformation to obtain the correct vector 
potential.  All in all, this is more work than our techniques.

\subsection{Cell-by-cell generation of the vector potential}
\label{cellbycell}

We present an algorithm for generating the vector
potential on a cell-by-cell basis that is linearly dependent on the
values of the magnetic field $\mathbf{B}$ across each face of the cell.
The divergence of $\mathbf{B}$ is assumed numerically zero; if
this isn't the case---say because data were interpolated---we recommend
use of a divergence cleaning method as a first step. Otherwise an
inconsistent vector potential will be produced, as we define our
staggered vector potential to produce a divergence-free
magnetic field by construction.

The basic outline of the cell-by-cell method is as follows:
\begin{itemize}
\item Choose an arbitrary point on the grid to be the origin, and use
  the ``six-face'' technique (Sec.~\ref{six-face}) to determine the
  $\mathbf{A}$-field values in that cell.  For all our tests, we started
  in the corner of the grid for which $i=j=k=0$.
\item Extend the solution along the six coordinate directions from the
  origin, using the ``five-face'' technique (Sec.~\ref{five-face}) to incorporate the four
  previously determined values for the overlapping edges of the shared
  face.
\item Further extend to the three ``coordinate planes,'' now using the ``four-face'' technique 
  (Sec.~\ref{four-face}) to take into account the seven previously computed edge values. 
  (Note: If the chosen origin is (0,0,0), these planes would be the $xy$-, $xz$-, and $yz$-planes)
\item Finish off the calculation by using the ``three-face'' technique  (Sec.~\ref{three-face})
  on all remaining cells in the grid.
\end{itemize}

In Table~\ref{table:numeq}, we list the number of times we use each of these techniques for a grid of size $L\times M\times N$, regardless of which cell is chosen to initiate the process.

\begin{table}[h!]
\centering
\begin{tabular}{|l|c|}
\hline
Method & \# of cells  \\ 
\hline
Six-face & $1$  \\
Five-face & $L+M+N-3$  \\
Four-face & $LM+LN+MN-2(L+M+N)+3$ \\
Three-face & $LMN-(LM+LN+MN)+(L+M+N)-1$ \\ 
\hline
\end{tabular}
\caption{Cell counts for each of the cell-by-cell techniques discussed below.}
\label{table:numeq}
\end{table}

\subsubsection{Initial cell: six undetermined faces}
\label{six-face}

For the first cell in the grid that we consider, there is no prior
information about the vector potential, and it must be completely
determined based on the six values of the magnetic field across the
faces.  The problem is under-determined without making some
assumptions. By assuming that whatever formulae we use to
compute the vector potential will be invariant under rotations of the
cube, we obtain a unique solution (as detailed in \ref{A:cell-by-cell}):
\begin{align}
A^i_{\tinysub{--}} &= \frac{5}{24}(-KB^j_{\tinysub{-}} + JB^k_{\tinysub{-}}) + \frac{1}{24}(-KB^j_{\tinysub{+}} + JB^k_{\tinysub{+}}) \, , \nonumber \\
A^i_{\tinysub{+-}} &= \frac{5}{24}(-KB^j_{\tinysub{+}} - JB^k_{\tinysub{-}}) + \frac{1}{24}(-KB^j_{\tinysub{-}} - JB^k_{\tinysub{+}}) \, , \nonumber \\
A^i_{\tinysub{++}} &= \frac{5}{24}( KB^j_{\tinysub{+}} - JB^k_{\tinysub{+}}) + \frac{1}{24}( KB^j_{\tinysub{-}} - JB^k_{\tinysub{-}}) \, , \nonumber \\
A^i_{\tinysub{-+}} &= \frac{5}{24}( KB^j_{\tinysub{-}} + JB^k_{\tinysub{+}}) + \frac{1}{24}( KB^j_{\tinysub{+}} + JB^k_{\tinysub{-}}) \, ,
\label{eq:6face}
\end{align}
where we may take any positive cyclic permutations of $x,y,z$ in terms
of $i,j,k$.

\subsubsection{Five undetermined faces}
\label{five-face}

Once the vector potential is calculated for an initial cell, we may
extend the solution to neighboring cells under the condition that the
vector potential values shared with previously constructed cells
are not changed.  

As the solution is propagated in each coordinate
direction from the initial cell, there will be cases in which one face and
the four edges that surround it have already been determined, leaving
five faces and eight edges yet to be determined. 
Call the previously determined face $B^i_-$, having denoted it $B^i_+$
when calculating vector potentials for the prior cell.  For the new
cell, we first compute a solution using the six-face technique
described in Sec.~\ref{six-face}, denoting it as $\tilde{A}$.  In general
this will lead to conflicting results for the previously determined
overlapping edges because nothing enforces consistency between
$\tilde{A}$ on our new cell compared to values from previous cells.
To remove inconsistencies, we propagate changes to the
non-overlapping, heretofore unspecified edges.  To do so, we add a
constraint based on consistency.  If the six-face solution $\tilde{A}$
is consistent with previously computed values, we adopt it and move
on. If not, we modify it on the eight undefined edges as follows.

Define $\hat{A}^j_{\pm -}$ and $A^k_{-\pm}$ to be the vector potential
values computed previously for the face encompassing $B^i_-$, and
define mismatches
\begin{equation} 
\delta{A}^j_{\pm-} \equiv \tilde{A}^j_{\pm-} - \hat{A}^j_{\pm-} \, ; ~~ \delta{A}^k_{-\pm } \equiv \tilde{A}^k_{-\pm } - \hat{A}^k_{-\pm } \, .
\label{eq:diffs}
\end{equation}
There are two consistent ways to determine the new edge values in a
symmetric way
(i.e. a way that is not preferential towards any one direction $x$, $y$, or $z$).
The simplest is to
propagate the differences from the edges surrounding the face $B^i_-$
to those surrounding the face $B^i_+$, making sure to keep the
orientation of the change correct, setting
\begin{align}
A^j_{\pm+} &= \tilde{A}^j_{\pm+} - \delta A^j_{\pm-} \, , \nonumber\\
A^k_{+\pm} &= \tilde{A}^k_{+\pm} - \delta A^k_{-\pm} \, , \nonumber \\
A^i_{\pm\pm} &= \tilde{A}^i_{\pm\pm} \, .
\label{eq:5face}
\end{align}
Alternately, one may change the values of the terms $A^i_{\pm\pm}$ on
the adjoining edges to the known face, leaving the edge values of the
opposing face unchanged.  We note that any linear combination of the two methods will also
generate a consistent vector potential for the new cell as well.  All
results shown here use only the first of these methods.

\subsubsection{Four undetermined faces}
\label{four-face}

Once the solution for the vector potential has been extended in each
cardinal direction from the initial cell, the next step is to expand
it into the coordinate planes, the planes for which one index is equal 
to its ``origin'' value.  If the chosen origin is (0,0,0), these planes
would be the $xy$-, $xz$-, and $yz$-planes.  This requires solving for the vector
potential for cells in which edges for two adjoining faces have been
determined, leaving five leftover edges spanning parts of the four
remaining faces.  

Again, we begin construction of the remaining vector
potential values by using the six-face method to construct a set of
values denoted $\tilde{A}$.  If we assume that the faces containing
the quantities $B^i_{\tinysub{-}}$ and $B^j_{\tinysub{-}}$ were
previously determined, our task is to determine the new values for
$A^i_{\tinysub{+\pm}}$, $A^j_{\tinysub{\pm+}}$, and
$A^k_{\tinysub{++}}$.  The resulting problem is similar to the
alternate approach for five undetermined faces, and we find
\begin{align}
A^i_{\tinysub{+-}} &= \tilde{A}^i_{\tinysub{+-}}-\frac{3\left( \delta A^i_{\tinysub{--}}-\frac{I}{J}\delta A^j_{\tinysub{--}}+\frac{I}{K}\delta A^k_{\tinysub{-+}}\right)
                                              +\left( \delta A^i_{\tinysub{-+}}-\frac{I}{J}\delta A^j_{\tinysub{+-}}+\frac{I}{K}\delta A^k_{\tinysub{+-}}\right)}{8} \, , \nonumber \\
A^j_{\tinysub{-+}} &= \tilde{A}^j_{\tinysub{-+}}-\frac{3\left(-\frac{J}{I}\delta A^i_{\tinysub{--}}+\delta A^j_{\tinysub{--}}+\frac{J}{K}\delta A^k_{\tinysub{+-}}\right)
                                              +\left(-\frac{J}{I}\delta A^i_{\tinysub{-+}}+\delta A^j_{\tinysub{+-}}+\frac{J}{K}\delta A^k_{\tinysub{-+}}\right)}{8} \, , \nonumber \\
A^i_{\tinysub{++}} &= \tilde{A}^i_{\tinysub{++}}-\frac{3\left( \delta A^i_{\tinysub{-+}}-\frac{I}{J}\delta A^j_{\tinysub{+-}}-\frac{I}{K}\delta A^k_{\tinysub{-+}}\right)
                                              +\left( \delta A^i_{\tinysub{--}}-\frac{I}{J}\delta A^j_{\tinysub{--}}-\frac{I}{K}\delta A^k_{\tinysub{+-}}\right)}{8} \, , \nonumber \\
A^j_{\tinysub{++}} &= \tilde{A}^j_{\tinysub{++}}-\frac{3\left(-\frac{J}{I}\delta A^i_{\tinysub{-+}}+\delta A^j_{\tinysub{+-}}-\frac{J}{K}\delta A^k_{\tinysub{+-}}\right)
                                              +\left(-\frac{J}{I}\delta A^i_{\tinysub{--}}+\delta A^j_{\tinysub{--}}-\frac{J}{K}\delta A^k_{\tinysub{-+}}\right)}{8} \, , \nonumber \\
A^k_{\tinysub{++}} &= \tilde{A}^k_{\tinysub{++}} \, .
\label{eq:4face}
\end{align}
Notice $A^k_{\tinysub{++}}$ plays a slightly different role than the
others as the edge bordering a pair of faces opposite the two has
already been determined.

\subsubsection{Three undetermined faces}
\label{three-face}

The generic case describing the majority of the cells within our grid
is one for which the vector potential has been determined on the nine
edges encircling three adjoining faces of the cell.   This leaves
three undetermined edge values and three undetermined faces.  If we
assume that all data have been determined on the faces supplying the
values for $B^i_{\tinysub{-}}$, $B^j_{\tinysub{-}}$, and
$B^k_{\tinysub{-}}$, then the only remaining vector potential values
left to be determined would be $A^i_{\tinysub{++}}$,
$A^j_{\tinysub{++}}$, and $A^k_{\tinysub{++}}$.  Once we have
determined the value of any of these three, the other two values
follow by requiring $\mathbf{B}=\nabla\times\mathbf{A}$ for 
each face.  Once one of the three values is set, there is at most 
one undetermined edge on any face, so they can be easily calculated.

Using the same conventions as before, we may work out the values using
our initial-cell techniques, defining differences as in the previous
cases, and then evaluating the symmetric formulae
\begin{align}
A^i_{\tinysub{++}}&=\tilde{A}^i_{\tinysub{++}}-\frac{1}{3}\left(\delta A^i_{\tinysub{+-}}+\delta A^i_{\tinysub{-+}}-\frac{I}{J}\delta A^j_{\tinysub{+-}}-\frac{I}{K}\delta A^k_{\tinysub{-+}}\right) \, , \nonumber\\
A^j_{\tinysub{++}}&=\tilde{A}^j_{\tinysub{++}}-\frac{1}{3}\left(\delta A^j_{\tinysub{+-}}+\delta A^j_{\tinysub{-+}}-\frac{J}{K}\delta A^k_{\tinysub{+-}}-\frac{J}{I}\delta A^i_{\tinysub{-+}}\right) \, , \nonumber\\
A^k_{\tinysub{++}}&=\tilde{A}^k_{\tinysub{++}}-\frac{1}{3}\left(\delta A^k_{\tinysub{+-}}+\delta A^k_{\tinysub{-+}}-\frac{K}{I}\delta A^i_{\tinysub{+-}}-\frac{K}{J}\delta A^j_{-+}\right) \, .
\label{eq:3face}
\end{align}

\subsubsection{Conversion to Coulomb gauge and removal of noise}
\label{coulomb}

After applying this cell-by-cell algorithm, we find our solution to
possess a few undesirable properties.  First,
nothing in these techniques will guarantee that the vector potentials
we generate in a cell-by-cell way remain smooth, and indeed we find
that by propagating changes across each cell in turn, we accrue rather
large changes within a cell by the time we reach the edges of our
grid, with potentially large jumps in the vector potential values on a
cell-by-cell basis.  Furthermore, the vector potential is determined
in a gauge that seems arbitrary and has no convenient mathematical
description.  To minimize the first of these problems and eliminate
the second, we convert our result to the Coulomb gauge using a
convolution technique that is described below.

In general, our numerically constructed vector potential will not
satisfy the Coulomb gauge condition.  If we assume that we can
determine a field $\phi$ to perform a gauge transformation of the form
$\mathbf{A}_C=\mathbf{A}-\nabla\phi$, where $\mathbf{A}_C$ is our
desired Coulomb gauge solution, under the condition that
$\nabla\cdot\mathbf{A}_C=0$, we find
\begin{align}
0 = \nabla\cdot\mathbf{A}_C &= \nabla\cdot\mathbf{A} - \nabla^2\phi \nonumber \\
\nabla^2\phi &= \nabla\cdot\mathbf{A} \, .
\label{eq:psi_phi}
\end{align}

There are numerous ways to solve the resulting scalar Poisson
equation, but given the absence of well-defined boundary conditions,
we choose a Fast Fourier Transform (FFT)-based convolution method.
Analytically, under the assumption that $\phi \to 0$ as $r \to
\infty$, the solution to Eq.~\eqref{eq:psi_phi} is given by
\begin{equation}
\phi = -\frac{1}{4\pi}\iiint \frac{\left[\nabla\cdot\mathbf{A}\right](\mathbf{r}')}{|\mathbf{r}-\mathbf{r}'|} \ud^3 \mathbf{r}'
      = \mathcal{F}^{-1}\left\{\mathcal{F}\left[\nabla\cdot\mathbf{A}\right]\star\mathcal{F}\left[-1/(4\pi r)\right]\right\} \, .
\label{eq:convolution}
\end{equation}
Here, the integral is evaluated over the volume of the computational
domain, the symbols $\mathcal{F}$ and $\mathcal{F}^{-1}$ represent
forward and reverse Fourier transforms respectively, and the $\star$ operator
implies that the transformed arrays are multiplied
element-by-element to perform a convolution.  In practice, we do not
actually convolve with the function $-1/4\pi r$ because it
will not have zero Laplacian when discretized, yielding a solution
that contains a nontrivial divergence.  Instead, we numerically solve
the three-dimensional Laplace equation for a $\delta$-function source
and use this as a basis for our convolution kernel, as described in 
\ref{A:coulomb}.

\subsection{Global linear algebra}
\label{global_LA}

Perhaps the most straightforward method to construct a staggered
vector potential configuration corresponding to a given magnetic field
is to treat the problem as a large,  sparse linear algebra
problem.  

Let us first consider exactly how many equations need to be
solved. According to Eq.~\eqref{eq:ABsizes}, there are
$3LMN+2(LM+LN+MN)+(L+M+N)$ vector potential values that must be set.
Our direct cell-by-cell method yields a count for the number of
equations that must be devoted to enforcing consistency between
magnetic field values on cell faces and the vector potential values
spanning that face.  Noting that  each ``$n$''-face method provides $n-1$ consistency equations along 
with one that will automatically be satisfied by the divergence-free criterion, 
we find a total count of $2LMN+(LM+LN+MN)$ equations
required to enforce consistency after summing over the cell counts shown in Table~\ref{table:numeq}.
To construct a well-posed system, the remaining equations, numbering
$LMN+(LM+LN+MN)+L+M+N=(L+1)(M+1)(N+1)-1$ must be specified to choose a
particular gauge condition.  We note that this number corresponds to
the total number of grid vertices present, save one.

While any gauge condition expressible in linear form may be chosen, we
will describe how to implement the Coulomb gauge condition for the
sake of consistency with the cell-by-cell method.  For vertices on the
interior of the grid, this simply means implementing
Eq.~\eqref{eq:Coulomb}, while for the vertices on the boundary we must
make some assumption about the vector potential components that lie
outside of the computational domain. We choose a simple, nonsingular\footnote{We 
note that a ``copy''-type boundary condition, in which the normal
components of the vector potential are set to be equal to those
immediately inside the boundary, yields a singular linear system that
cannot be evaluated, as it permits a constant non-zero vector
potential solution for zero source.} option: zero
the normal components of the vector potential at the boundary of the
domain. 

While it is straightforward to construct a sparse matrix problem whose
solution is the desired vector potential configuration, 
computational efficiency concerns force us to consider the organization of
the linear system.  Unlike the case for calculating the convolution
kernel for Coulomb gauge conversion described in Sec.~\ref{coulomb},
this matrix cannot be made diagonally dominant,
nor can it easily be constructed in symmetric form, so several efficient
techniques for solving sparse systems are immediately ruled out.  We
have found that most solvers perform better when the
diagonal terms are non-zero in each row, and when the bandwidth of the
non-zero elements is minimized, which motivates our approach to the
problem.  In what follows, we discuss a straightforward approach to
generating a sparse linear system that yields a vector potential
solution in the Coulomb gauge, satisfying the principles above. 

Our linear system consists of one equation for each of the unknown
vector potential values: a total of $3LMN+2(LM+LN+MN)+L+M+N$
equations.  Each of the vector 
potential components is organized in dimension-by-dimension order,
with the $z$-coordinate varying most rapidly and the $x$-coordinate
the slowest.  In order to maximize geometric proximity of neighboring
elements in our linear system, we recommend interleaving the vector
potential components; thus, successive rows of our matrix represent
$A^x$, $A^y$, and $A^z$ components in turn (this is easiest for cubic
grids, for which the different vector potential components contain the
same number of elements on the grid).

The first set of linear equations is found by associating each $A^x$
value with a corresponding instance of the Coulomb gauge condition.
In particular, the matrix row corresponding to a value
$A^x(i,j+\onehalf,k+\onehalf)$ is associated with the equation
\begin{equation}
0 = [\nabla\cdot\mathbf{A}](i+\onehalf,j+\onehalf,k+\onehalf) \, ,
\label{eq:gla_Ax}
\end{equation}
where the right-hand side is evaluated using Eq.~\eqref{eq:Coulomb}.
Again, any vector potential value lying outside the computational
domain is set to zero.  This has the effect
of enforcing the Coulomb condition at every grid vertex in the domain
except those on the ``leftmost'' face with coordinates
$(-\onehalf,j+\onehalf,k+\onehalf)$, which are handled later.

For the rows corresponding to $A^y$ values, we use two different types
of equations.  For the leftmost set of values, i.e., those with
$i=-\onehalf$, the Coulomb condition is enforced, such that for the
row corresponding to $A^y(-\onehalf,j,k+\onehalf)$, our matrix row
implements the equation
\begin{equation}
0 = [\nabla\cdot\mathbf{A}](-\onehalf,j+\onehalf,k+\onehalf) \, ,
\label{eq:gla_Ay}
\end{equation}
with the same treatment as above for components lying outside the
computational domain.  Combined with the previous step, the only
vertices at which the Coulomb condition has not been applied are those
on the grid edge satisfying the condition $i=j=-\onehalf$.

For the remaining rows corresponding to $A^y$ values, we demand
consistency with the $B^z$ value for the given magnetic field.  In
particular, for the row corresponding to
$A^y(i+\onehalf,j,k+\onehalf)$, we implement the equation
\begin{align}
B^z(i,j,k+\onehalf) =& \frac{A^y(i+\onehalf,j,k+\onehalf)-A^y(i-\onehalf,j,k+\onehalf)}{I} \nonumber \\
  &- \frac{A^x(i,j+\onehalf,k+\onehalf)-A^x(i,j-\onehalf,k+\onehalf)}{J} \, .
\label{gla_AyBz}
\end{align}

For rows corresponding to $A^z$ values, three sets of
equations must be implemented:
\begin{enumerate}
\item For those on the edge of the
domain, with coordinates $i=j=-\onehalf$, the Coulomb
gauge condition is applied, so that for the row corresponding to
$A^z(-\onehalf,-\onehalf,k)$, we implement
\begin{equation}
0 = [\nabla\cdot\mathbf{A}](-\onehalf,-\onehalf,k+\onehalf) \, .
\label{eq:gla_Az}
\end{equation}
At this point, the Coulomb condition has been enforced at every vertex
within the domain, except the corner point with coordinates
$i=j=k=-\onehalf$.  This serves as the single point at which we do not
have the degrees of freedom available to implement the Coulomb
condition.

\item For remaining $A^z$ values with coordinates $i=-\onehalf$, we
enforce consistency for given values of $B^x$.  In particular, for
rows corresponding to the values $A^z(-\onehalf,j+\onehalf,k)$, we
implement
\begin{align}
B^x(-\onehalf,j,k) =& \frac{A^z(-\onehalf,j+\onehalf,k) - A^z(-\onehalf,j-\onehalf,k)}{J} \nonumber \\
               &- \frac{A^y(-\onehalf,j,k+\onehalf)-A^y(-\onehalf,j,k-\onehalf)}{K} \, .
\label{eq:gla_AzBx}
\end{align}

\item For the remaining rows corresponding to $A^z$ values lying
elsewhere, with $i\ne -\onehalf$, we enforce consistency for the given
$B^y$ values.  Rows corresponding to values
$A^z(i+\onehalf,j+\onehalf,k)$ are used to solve the equations
\begin{align}
B^y(i,j+\onehalf,k) =& \frac{A^x(i,j+\onehalf,k+\onehalf)-A^x(i,j+\onehalf,k-\onehalf)}{K} \nonumber \\
                &-\frac{A^z(i+\onehalf,j+\onehalf,k) -A^z(i-\onehalf,j+\onehalf,k)}{I} \, .
\label{eq:gla_AzBy}
\end{align}
\end{enumerate}

For input magnetic field configurations that are  divergence-free, our method yields the unique vector potential consistent with both it as well as the gauge and boundary conditions.  For a magnetic field that {\em does contain numerical divergences}, our method acts as an implicit one-dimensional ``divergence cleaner'', sweeping numerical divergences off of the grid in  a cell-by-cell fashion, working from the $B^x$ values on the leftmost face of the grid and transferring away divergences to the right and eventually out of the rightmost face of the grid.
Our choice of the roles of the $x$-,
$y$-, and $z$-directions is arbitrary, chosen simply for convenience.

The linear system of equations was implemented in the \mumps
\cite{Amestoy:2001,Amestoy:2006,MUMPS:web} infrastructure, an MPI-parallelized
sparse matrix package. With it, we are capable of generating suitable
vector potentials even on numerical grids of order $100^3$
in size, as discussed and demonstrated in the following sections.

\section{Code performance}
\label{performance}

We have tested the performance of our code on the \NH computational
cluster at RIT.  Overall, it consists of 64-bit AMD and Intel CPUs 
interconnected with a high-speed, low-latency QDR InfiniBand fabric,
and 4 GB of RAM per core.  Specifically, our simulations were run on 
a subset of \NH with dual-core AMD Opteron$^{\mathrm{TM}}$ processors.

\begin{figure}[h!]
\centering
\includegraphics[width=3.5in]{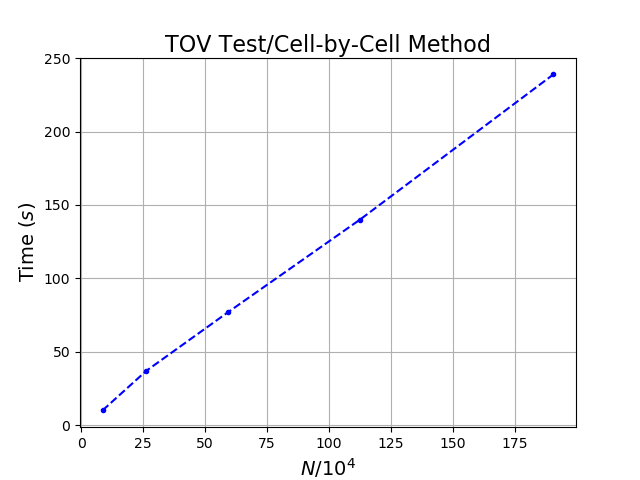}
\caption{Scaling results for the cell-by-cell method, as measured in
  walltime versus number of grid cells.}
\label{timing_CBC}
\end{figure}

Our cell-by-cell method is written as a serial
code. As shown in Fig. \ref{timing_CBC}, the computational time it
requires scales linearly with the number of grid cells, and can be
expected to follow this behavior even for much larger
grids.  Because the cell-by-cell method is strictly serial, it must 
be run on a single core, so the CPU core time is equal to the walltime.  
The code converges for all resolutions on a single core.  
It requires very little memory to generate its initial vector
potential configuration beyond that required to store the magnetic
field and vector potential values. Larger grids are required to
perform the FFT convolution while converting the vector potential to
the Coulomb gauge, but if need be this step could be separated from
the rest of the code and run in parallel using the existing FFTW
infrastructure \cite{Frigo:2005}, which is the leading
MPI-parallelized multi-dimensional FFT package.
 
\begin{figure}[h!]
\centering
\includegraphics[width=3.5in]{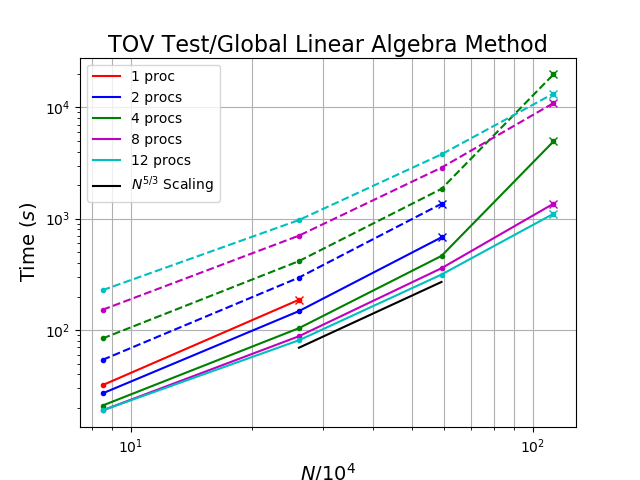}
\caption{Walltime and CPU core time for the global linear algebra method
  run on different numbers of cores on \NH. For a given number of
  CPU cores, the solid line represents the walltime, and the dashed
  line the CPU core time.  Points marked with $\times$'s correspond to the
  highest resolution $N$ for which the code converges with the given
  amount of RAM (4GB/core on \NH).}
\label{timing_GLA}
\end{figure}

Fig. \ref{timing_GLA} shows that the walltime required to use our
global linear algebra method largely scales as the number
of gridpoints, $N$, to the $\sfrac{5}{3}$ power, as expected due to 
how \mumps implements basic Block-Low Rank (BLR) factorization 
\cite{Amestoy:2017}.  The only significant deviation from this pattern 
appears at large grid sizes, for cases in which RAM runs out and 
swapping occurs. More minor deviations from this pattern occur due to 
matrix inversion requiring a significant communication overhead (handled
internally by the \mumps package).  Thus the use of additional cores
typically increases the required CPU core time to complete a run, even if
walltime is reduced.  In terms of memory, this method scales as $N$ to
the \sfrac{4}{3} power.

\section{Numerical tests}
\label{tests}

To test our codes, we use the \IGM\footnote{See \ref{A:ghost} for a discussion of how ghost
cells are treated in the context of interfacing our methods with
\IGM.} code within the \ET as our dynamical evolution
code for various magnetized fluid configurations, as it evolves the
magnetic vector potential over time as its dynamical variable.  For
each configuration, we perform the simulation twice.  

First, we run \IGM to the designated final time as a baseline. We call
this the ``uninterrupted run''.  Next we start a second, ``restarted'' run, in which the simulation
proceeds to a predetermined ``checkpoint'' time, typically halfway
through the simulation.  At this point, we output the $\mathbf{A}$ field
and numerically compute a $\mathbf{B}$ field from it (via the definition of $\mathbf{A}$)
and use these magnetic fields as input for our cell-by-cell and
global-linear-algebra codes. We then generate a new vector potential
$\mathbf{A}'$ from the $\mathbf{B}$ field, which will typically differ
significantly from those generated by \IGM. We feed $\mathbf{A}'$ back
into \IGM and restart the run at the checkpoint time and continue
until the final time. 

We expect that if our approach is valid, the final {\it magnetic
  fields} (i.e., the physical, as opposed to the gauge-dependent,
field) will agree to many significant digits between the restarted or
uninterrupted runs. What follows is a demonstration of the validity of
our approach in a variety of contexts: a magnetized two-dimensional,
relativistically-spinning rotor (Sec.~\ref{Rotor}), a
weakly-magnetized neutron star (Sec.~\ref{TOVstable}), and an
extremely-magnetized neutron star (Sec.~\ref{TOVunstable}).

\subsection{Rotor test}
\label{Rotor}

The rotor test, as described in \cite{Mosta:2014} consists of a disk
of material in the $xy$-plane, referred to as a rotor, which possesses
ten times the density of the surrounding material.  The rotor is
spinning such that its edge is moving at $0.995c$, and the entire
medium is permeated by a uniform magnetic field in the $x$-direction.
We run the simulations until the rotor has undergone two-thirds of a
full rotation, with a checkpoint time ($t_c$) of one-third a full
rotation.

\begin{figure}[h!]
\centering
\includegraphics[width=3.5in]{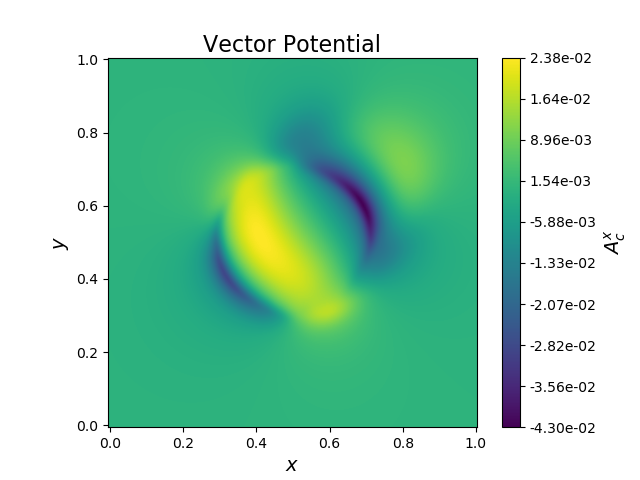}
\caption{The magnetic vector potential $A^x$ plotted vs $x$ and $y$ at 
the checkpoint time $t_c$ after running through the cell-by-cell method.  
All quantities are in arbitrary units.}
\label{fig:Rotor_CBC_Ax_xy_320}
\end{figure}

\begin{figure}[h!]
\centering
\includegraphics[width=3.5in]{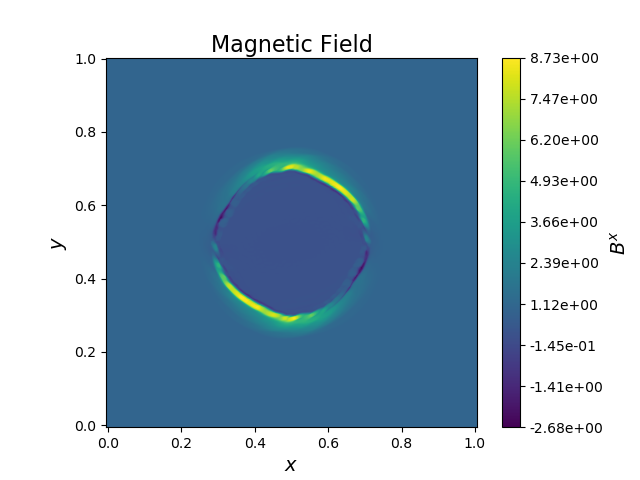}
\caption{The magnetic field $B^x$ plotted vs $x$ and $y$ at the checkpoint 
time $t_c$ after running through the cell-by-cell method.}
\label{fig:Rotor_CBC_Bx_xy_320}
\end{figure}

Fig. \ref{fig:Rotor_CBC_Ax_xy_320} shows one component of the magnetic
vector potential, $A^x$, in arbitrary units, plotted in the $xy$-plane at the checkpoint
time.  These are data from the output of the cell-by-cell method.
Fig. \ref{fig:Rotor_CBC_Bx_xy_320} displays the magnetic field
component $B^x$ on the same plane at the same time, also from the
output of the cell-by-cell code.  By inspection, these fields appear
fairly smooth in their own right, but are they consistent with those
of the uninterrupted run?

\begin{figure}[h!]
\centering
\includegraphics[width=3.5in]{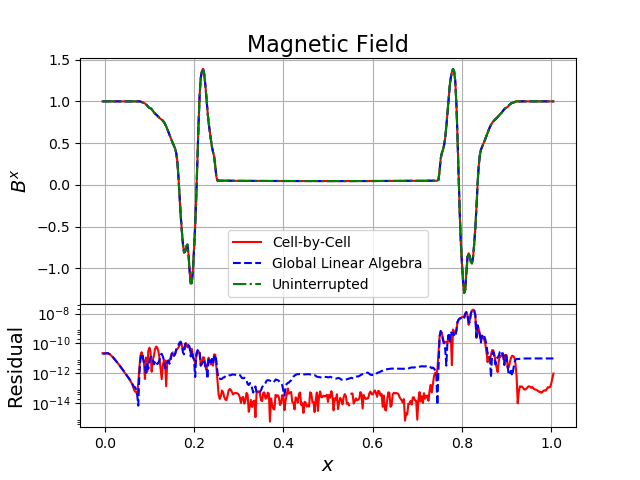}
\caption{\textit{Top:} The magnetic field $B^x$ vs $x$ at the final time 
$t_f$ after running the rotor through the cell-by-cell and global linear algebra methods.  
\textit{Bottom:} The absolute difference between the cell-by-cell and global 
linear algebra runs and the corresponding uninterrupted run.}
\label{fig:Rotor_Bx_x_BOTH_640}
\end{figure}

To compare the two runs of \IGM, we look at the components of the
magnetic field $\mathbf{B}$.  We do not directly compare the vector
potential $\mathbf{A}$ between the two runs because the chosen vector
potential gauges are different, depending both on whether an inverse
curl algorithm is applied and on which such algorithm is chosen. Thus
the final $\mathbf{A}$ field can be very different between uninterrupted
and restarted runs, yet still yield the same magnetic field. This is
why we say the magnetic field is the physically relevant quantity.
Fig. \ref{fig:Rotor_Bx_x_BOTH_640} displays the dominant
magnetic-field component, $B^x$, versus $x$ at the final
time for all three runs: uninterrupted, restarted with cell-by-cell data, and
restarted with global linear algebra data. Notice the agreement is within about
one part in $10^8$ or more throughout the entire data set. 

\subsection{Magnetized neutron stars}
\label{TOV}

Here we consider models of neutron stars with interior, poloidal
magnetic fields, described using the Tolman-Oppenheimer-Volkhoff (TOV)
model, a configuration that is also described in \cite{Mosta:2014}.
In Sec.~\ref{TOVstable}, we start with a magnetic field strength parameter of $A_b \approx
0.64$, which results in a ratio of magnetic pressure to gas pressure
of $b^2/2P \sim 0.001$.  This should result in a star with
little to no evolution due to magnetic effects, so we call this the
``stable'' case.  Then, in Sec.~\ref{TOVunstable}, we set the initial
magnetic field strength a value $10^4$ times larger, resulting in a
magnetic to gas pressure ratio $10^8$ times larger.  This should
cause the magnetic pressure to violently blow apart the star, so we
refer to this as the ``unstable'' case.  The timescale of interest in
these tests is the dynamical timescale of 
the stable TOV star, $\tdyn$.  The final time for all tests is fixed
at $4\,\tdyn$, and the checkpoint time is set to $\tdyn$.

\subsubsection{Magnetically stable case}
\label{TOVstable}

\begin{figure}[h!]
\centering
\includegraphics[width=3.5in]{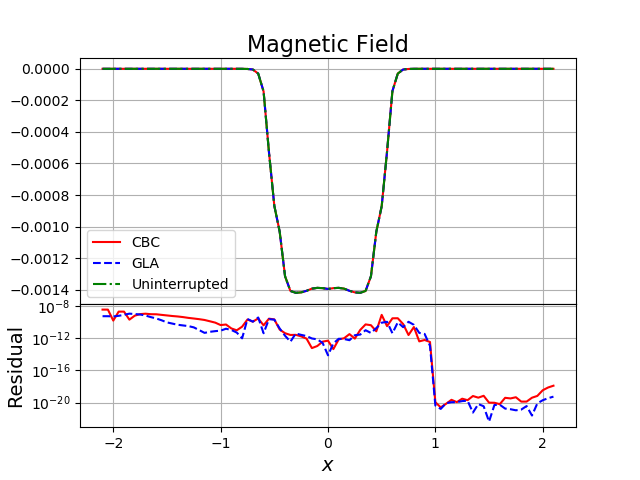}
\caption{\textit{Top:} The resultant $B^z$ vs $x$ after running the stable 
TOV star through the cell-by-cell and global linear algebra routines and 
restarting. \textit{Bottom:} The absolute difference between the cell-by-cell 
and global linear algebra runs and the corresponding uninterrupted run.  The 
checkpoint time was the dynamical time, and the final time was four times the dynamical time.}
\label{fig:TOV_Bz_x_S_BOTH_td1}
\end{figure}

As expected, there was not much evolution of the magnetic,
hydrodynamic, or gravitational fields in the ``stable''
configuration. The neutron star slowly relaxes from its initial
state.  Fig. \ref{fig:TOV_Bz_x_S_BOTH_td1} shows the dominant magnetic
field component ($B^z$) along the $x$ axis for the cell-by-cell and global linear algebra runs at the
final time $4\,\tdyn$.  Conventions are the same as in Fig. \ref{fig:Rotor_Bx_x_BOTH_640}.  
Agreement between runs is exceptional, particularly in the region of
interest: the interior of the star ($|x|\lesssim 0.65$) where the
magnetic fields remain nonzero throughout the evolution. 

However, small disagreements exist at the boundary, due to the
hybrid quadratic-cubic boundary conditions described in \ref{A:ghost}.
These boundary conditions pose some difficulty for numerical evolutions, as
even the uninterrupted case shows the development of magnetic
phenomena near the grid boundary.  Also note this effect is only present on one side of 
the grid, when $x\approx -2$.  We believe this to be due to \IGM's extrapolation of
$\mathbf{A}$ to fill in ghost cells at the boundary.  In a full-scale
simulation, the boundary will be sufficiently far away that such edge
effects will not be a problem.

\subsubsection{Magnetically unstable case}
\label{TOVunstable}

\begin{figure}[h!]
\centering
\includegraphics[width=3.5in]{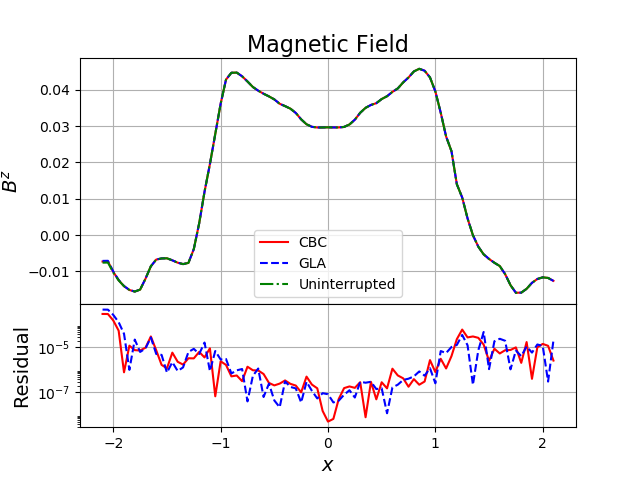}
\caption{\textit{Top:} The resultant $B^z$ vs $x$ after running the unstable 
TOV star through the cell-by-cell and global linear algebra routines and 
restarting. \textit{Bottom:} The absolute difference between the cell-by-cell 
and global linear algebra runs and the corresponding uninterrupted run.  The 
checkpoint time was the stable dynamical time, and the final time was four times the stable dynamical time.}
\label{fig:TOV_Bz_x_U_BOTH_td1}
\end{figure}

Our inverse curl algorithms also performed well in this test, though
agreement dropped to about one part in $10^4$, likely due to the
transmission of errors from the boundary (the simulation lasts
$\approx 9$ light-crossing times). Over
the course of the simulation, the star blew itself apart, ejecting
material towards the edge of the grid.  The data for $B^z$ vs $x$ are
displayed in Fig. \ref{fig:TOV_Bz_x_U_BOTH_td1} 
using the same conventions as Fig. \ref{fig:TOV_Bz_x_S_BOTH_td1} in the previous section.  
Here the data on both ends of the grid ($x\approx -2$ and $x\approx 2$) actually 
match each other better than in the stable case.  We think this is due to the 
behavior seen in the stable case being wiped out by
the error caused by the outflow of the star exploding.

\section{Conclusion}
\label{conclusion}

The techniques described here are sufficient to construct a vector
potential on a staggered numerical grid, appropriate for use in
applications in which magnetic fields evolved using, e.g., a
constrained transport approach, must be converted into a vector
potential.  Our solution consists of a numerical
implementation of the ``inverse curl'' operator over a finite
rectangular domain.  This problem is typically much more difficult to
solve than the case of a spherical domain, for which a spectral
solution can be written down explicitly in terms of vector spherical
harmonic modes.  Additionally, these methods can be applied to any
problem in which two quantities are related by the curl, such as 
the fluid velocity and vorticity in the context of fluid dynamics.

Each of our techniques possesses unique strengths and weaknesses.  The
cell-by-cell method is very fast and scales as $\mathcal{O}(N)$ in
both time and memory.  On the other hand, the global linear algebra
method requires more memory and is slower, but it is a much more
symmetric technique that uniformly cleans nonzero divergences in the
input magnetic fields and applies Coulomb gauge
conditions as it solves, instead of as a separate step.  The global
linear algebra method is also much more amenable to the implementation
of different boundary conditions and mesh refinement than the
cell-by-cell method.  

Mesh refinement is an important tool used in numerical relativity
simulations to apply additional resolution in regions where it is necessary,
such as near black holes or neutron stars.  Both \harm and \IGM have
the capability to use mesh refinement, and so any method aimed at
bridging the gap between them via inverse curl algorithms must
also have such capabilities.  Unfortunately, the staggering of the
grid in the cell-by-cell method makes mesh refinement, even by a
simple factor of two, prohibitively complicated and most likely
infeasible.  The global linear algebra method, however, can handle
mesh refinement with careful consideration.  Another way to implement
mesh refinement is to use an inherently multi-grid method, which we
explore in a upcoming paper.

\section*{Acknowledgments}

Z.J.S. and J.A.F. were supported in part by NSF award OAC-1550436.  
Computational resources were provided by the \NH cluster at
RIT, which was supported by NSF grants No. PHY-0722703, 
No. DMS-0820923, and No. AST-1028087.
Support at WVU was provided in part by NSF EPSCoR Grant
OIA-1458952. Some early \mumps computations were performed on WVU's
\texttt{Spruce Knob} high-performance computing 
cluster, funded in part by NSF EPSCoR Research Infrastructure
Improvement Cooperative Agreement \#1003907, the state of West
Virginia (WVEPSCoR via the Higher Education Policy Commission), and
West Virginia University.
The authors would like to thank M. Avara, D. Bowen, M. Campanelli, 
B. Ireland, J. Krolik, C. Lousto, V. Mewes, S. Noble, J. Schnittman, 
and Y. Zlochower for helpful conversations.

\appendix

\section{Notes on the direct cell-by-cell solution method}
\label{A:cell-by-cell}

\subsection{Six faces}
\label{A:6faces}

\begin{figure}[h!]
\centering
\includegraphics[width=4.0in]{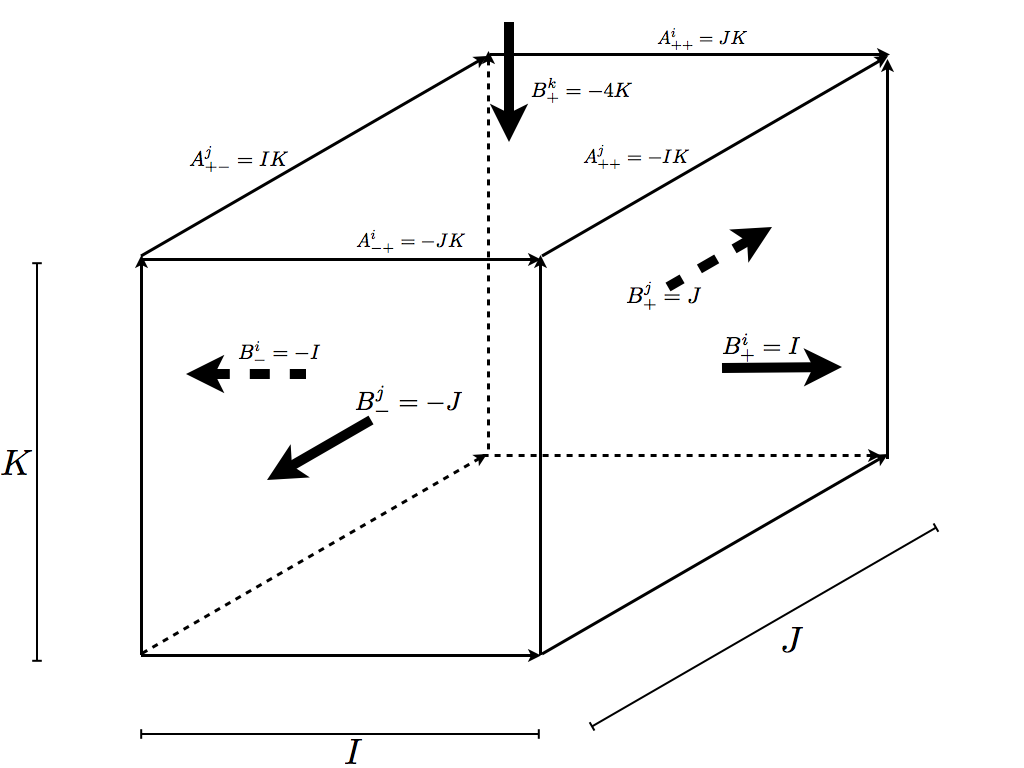}
\caption{The sample configuration used to calculate $\alpha$ and
  $\beta$.  This configuration is a magnetic field that enters the top
  of the box and uniformly leaves the sides.  It results in the vector
  potential circulating around the four edges composing the top face,
  with the other eight vector potential values all evaluating to
  zero.} 
\label{fig:config}
\end{figure}

To derive the ``six-face'' expressions for the vector potential
values, a single configuration can be constructed to establish a
maximally symmetric choice for the coefficients (i.e. a choice 
that is not preferential towards any one direction $x$, $y$, or $z$).
In Fig. \ref{fig:config}, a particular magnetic field
configuration is shown, in which a magnetic field of magnitude
$B^k_{\tinysub{+}}=-4K$ can be seen as entering a cell from the top,
while the values $B^i_{\tinysub{\pm}}=\pm I,~B^j_{\tinysub{\pm}}=\pm
J$ represent the same field leaving the cell equally through the four
sides.  No flux enters or leaves through the bottom of the cell, where
we set $B^k_{\tinysub{-}}=0$.  For this configuration, symmetry
arguments can be used to specify each of the vector potential values.
Around the top face of the cell, we expect the vector potential values
to all be equal (up to a common scaling factor) depending on the
dimensions of the cell:
\begin{equation}
A^i_{\tinysub{++}} = -A^i_{\tinysub{-+}} = JK \, ; ~~ A^j_{\tinysub{++}} = -A^j_{\tinysub{+-}} = -IK \, ,
\label{eq:configtop}
\end{equation}
while those on the bottom should all be zero:
\begin{equation}
A^i_{\tinysub{\pm-}} = A^j_{\tinysub{-\pm}} = 0 \, .
\label{eq:configbot}
\end{equation}
If we assume, based on the symmetry of the problem, that the vector
potential on a given edge picks up a contribution proportional to a
quantity $\alpha$ for the two faces that border the edge and a
contribution $\beta$ from the two on the opposite sides from these, we
find, for an edge value on the top face, that
\begin{align}
A^i_{\tinysub{++}} &= \alpha (KB^j_{\tinysub{+}} - JB^k_{\tinysub{+}}) + \beta (KB^j_{\tinysub{-}} - JB^k_{\tinysub{-}}) \, , \nonumber \\
JK &= \alpha (5JK) - \beta (JK) ~~ \rightarrow ~~ 5\alpha - \beta = 1 \, ,
\label{eq:topedge}
\end{align}
while for one on the bottom face,
\begin{align}
A^i_{\tinysub{--}} &= \alpha (-KB^j_{\tinysub{-}} + JB^k_{\tinysub{+}}) + \beta (-KB^j_{\tinysub{+}} + JB^k_{\tinysub{+}}) \, , \nonumber \\
0 &= \alpha (JK) + \beta (-5JK) ~~ \rightarrow ~~ \alpha- 5\beta = 0 \, .
\label{eq:botedge}
\end{align}
The solution to this set of equations is $\alpha=5/24$, $\beta=1/24$. 

This solution implies that for this configuration, the vertical edges,
with values $A^k_{\pm\pm}$, which must be equal but could be assigned
any value to yield a self-consistent magnetic field, will be set to
zero.  This seems appropriate, as it implies that for a spatially
uniform magnetic field oriented along one of the coordinate axes, the
vector potential values will be purely normal with no parallel
component.

\subsection{Five faces}
\label{A:5faces}

When calculating the vector potentials for the five-face method, there
are two independent methods by which differences may be symmetrically
propagated between the previously calculated solution and the new
six-face solution from the predetermined edges to those yet to be
calculated.  These differences may be computed in at least the
following two ways:
\begin{itemize}
\item Propagate the differences from the face $B^i_{\tinysub{-}}$ to
  $B^i_{\tinysub{+}}$, making sure to keep the orientation of the
  change correct, using the result from Eq.~\eqref{eq:5face}.
\item Change only the values of the terms $A^i_{\tinysub{\pm\pm}}$ on
  the adjoining edges to the known face, leaving the edge values of
  the opposing face unchanged.  The symmetric form of this operation
  is given by the set of equations:
\begin{align}
A^i_{\tinysub{++}} &= \tilde{A}_{\tinysub{++}}-\frac{I\left[3(-K\delta A^j_{\tinysub{+-}}-J\delta A^k_{\tinysub{-+}})+(-K\delta A^j_{\tinysub{--}}-J\delta A^k_{\tinysub{--}})\right]}{8JK} \, , \nonumber \\
A^i_{\tinysub{+-}} &= \tilde{A}_{\tinysub{-+}}-\frac{I\left[3(-K\delta A^j_{\tinysub{--}}+J\delta A^k_{\tinysub{-+}})+(-K\delta A^j_{\tinysub{+-}}+J\delta A^k_{\tinysub{--}})\right]}{8JK} \, , \nonumber \\
A^i_{\tinysub{--}} &= \tilde{A}_{\tinysub{--}}-\frac{I\left[3( K\delta A^j_{\tinysub{--}}+J\delta A^k_{\tinysub{--}})+( K\delta A^j_{\tinysub{+-}}+J\delta A^k_{\tinysub{-+}})\right]}{8JK} \, , \nonumber \\
A^i_{\tinysub{-+}} &= \tilde{A}_{\tinysub{+-}}-\frac{I\left[3( K\delta A^j_{\tinysub{+-}}-J\delta A^k_{\tinysub{--}})+( K\delta A^j_{\tinysub{--}}-J\delta A^k_{\tinysub{-+}})\right]}{8JK} \, , \nonumber \\
A^j_{\tinysub{\pm+}} &= \tilde{A}^j_{\tinysub{\pm+}} \, , \nonumber \\
A^k_{\tinysub{+\pm}} &= \tilde{A}^k_{\tinysub{+\pm}}\, ,
\label{eq:adjoin}
\end{align}
where the differences in the equations above are subject to a consistency condition:
\begin{equation}
J(\delta A^k_{\tinysub{-+}} - \delta A^k_{\tinysub{--}}) - K(\delta A^j_{\tinysub{+-}} - \delta A^j_{\tinysub{--}}) = 0 \, .
\label{eq:consist}
\end{equation}
\end{itemize}
We note that any linear combination of the two methods described above will also yield a valid solution.

\section{Convolution technique for Coulomb gauge condition}
\label{A:coulomb}

In what follows, we will simplify our discussion by assuming that our 
physical grid is cubic, with $N$ points in each dimension labeled from 
1 to $N$, with  uniform grid spacing $I$ in each of the three directions.  
We note, though, that these methods may be trivially generalized for cases 
where either the grid dimensions, grid spacings, or both are different 
in different directions.

In order to solve Eq.~\eqref{eq:convolution} via a convolution method, 
we will require a solution $\chi$ to the equation
\begin{equation}
\nabla^2 \chi = \delta(r) \, ,
\label{eq:chi0}
\end{equation}
which will serve as a convolution kernel.  This kernel function must be 
evaluated on a grid at least one cell larger in each direction than our 
physical grid, indexed so that the origin is placed at a corner and the 
spatial domain is thought of as an octant.  Thus we assume it is of size 
$N_{\rm ker}^3$, where $N_{\rm ker}\ge N+1$ and the indices run from 0 to 
$N_{\rm ker}-1$.  Ideally, one should choose the value of $N_{\rm ker}$ to be 
the smallest allowed value that can be expressed in the form $2^m+1$, for 
positive integer $m$, or a similar quantity involving small-integer 
factors, in order to perform the FFT described below.

The analytic solution for Eq.~\eqref{eq:chi0} is well known to be 
$\chi=-1/4\pi r$, where $r$ is the three-dimensional distance 
from a point to the origin, $r(i,j,k)=I\sqrt{i^2+j^2+k^2}$, but this 
function $\chi$ is singular at the origin and not an exact solution if 
we interpret the Laplacian operator as a finite-differencing expression.  
Instead, we can solve
\begin{align}
\chi^*(i-1,j,k) &+ \chi'(i+1,j,k) + \chi^*(i,j-1,k) + \chi'(i,j+1,k) \nonumber \\
                &+ \chi^*(i,j,k-1) + \chi'(i,j,k+1) - 6\chi(i,j,k) = \begin{cases}I^2; & i=j=k=0 \\ 
                                                         0; & {\rm otherwise}\end{cases} \, ,
\label{eq:findiff_chi}
\end{align}
where the notation $\chi^*$ is used to describe the appropriate boundary 
conditions that we will apply at the three faces of the cube describing 
coordinate planes, and the notation $\chi'$ is used to describe those that 
we will apply at the three faces that lie on the exterior.  At the former 
faces, we impose reflection symmetry, such that
\begin{equation}
\chi^*(i,j,k) = \begin{cases}\chi^*(1,j,k); & i=-1 \\
                                          \chi^*(i,1,k); & j=-1 \\
                                          \chi^*(i,j,1); & k=-1 \\
                                             \chi(i,j,k);& {\rm otherwise}\end{cases} \, ,
\label{eq:chistar}
\end{equation}
in order to handle cases where a neighboring value lies outside the grid 
across one of the coordinate planes.  At our outer boundaries, we  impose 
a $1/r$ falloff condition:
\begin{equation}
\chi'(i,j,k) = \begin{cases}\frac{r(N_{\rm ker}-1,j,k)}{r(N_{\rm ker},j,k)} \chi'(N_{\rm ker}-1,j,k); & i=N_{\rm ker} \\ \\
                                         \frac{r(i,N_{\rm ker}-1,k)}{r(i,N_{\rm ker},k)} \chi'(i,N_{\rm ker}-1,k); & j=N_{\rm ker} \\ \\
                                         \frac{r(i,j,N_{\rm ker}-1)}{r(i,j,N_{\rm ker})} \chi'(i,j,N_{\rm ker}-1); & k=N_{\rm ker} \\ \\
                                                                                               \chi(i,j,k); & {\rm otherwise}\end{cases} \, ,
\label{eq:chiprime}
\end{equation}
though the Dirichlet condition $\chi'(i,j,k)=0$ would also be a valid 
choice.  The resulting linear system, which is sparse and diagonally 
dominant, can be solved using any standard linear algebra package.  We 
have done so using \eigen \cite{eigen:web}.

To perform the FFT-based convolution, we need to map the convolution kernel 
function and the grid containing the divergence of the vector potential into 
a larger grid, to avoid aliasing effects.  Assuming that we have chosen the 
value of $N_{\rm ker}$ appropriately, our FFT grids will have dimensions 
$N_{\rm fft}\times N_{\rm fft}\times N_{\rm fft}$, where $N_{\rm fft} = 2(N_{\rm ker}-1)$ 
and we assume the indices range from 0 to $N_{\rm fft}-1$ in each direction.  
The convolution kernel function is 
\begin{equation}
\chi_{\rm fft}(i,j,k) = \chi(I,J,K) \, ,
\label{eq:chifft}
\end{equation}
where $I =\min(i,N_{\rm fft}-i)$, $J=\min(j,N_{\rm fft}-j)$, and $K=\min(k,N_{\rm fft}-k)$.  
The data array is given by
\begin{equation*}
\psi_{\rm fft}(i,j,k) = \begin{cases}\left[\nabla\cdot\mathbf{A}\right](i-\onehalf,j-\onehalf,k-\onehalf); & i<N~{\rm and}~j<N~{\rm and}~k<N \\
                                                                      0; & {\rm otherwise}\end{cases} \, .
\label{eq:psifft}
\end{equation*}
We may now evaluate
\begin{equation}
\phi_{\rm fft} =  \mathcal{F}^{-1}\left\{\mathcal{F}[\psi_{\rm fft}]\star\mathcal{F}[\chi_{\rm fft}]\right\} \, ,
\label{eq:phifft}
\end{equation}
where the symbols $\mathcal{F}$ and $\mathcal{F}^{-1}$ represent forward and 
reverse Fourier transforms, and the $\star$ operator implies that we multiply 
the transformed arrays against each other element by element.  Our implementation 
uses the {\tt FFTW} package to perform these operations.  Having calculated this 
field, we may then set
\begin{equation}
\phi(i-\onehalf,j-\onehalf,k-\onehalf) = \phi_{\rm fft}(i,j,k) \, ,
\label{eq:phi_phifft}
\end{equation}
for all indices $1\le i\le N$, $1\le j\le N$, and $1\le k\le N$ 
and then set $\mathbf{A}_c=\mathbf{A}-\nabla\phi$ using our staggered coordinates, e.g., 
\begin{align}
A_c^x(i,j+\onehalf,k+\onehalf) =& A^x(i,j+\onehalf,k+\onehalf) \nonumber \\
                      &- \frac{\phi(i+\onehalf,j+\onehalf,k+\onehalf) - \phi(i-\onehalf,j+\onehalf,k+\onehalf)}{I} \, .
\label{eq:Acx}
\end{align}
The resulting vector potential configuration satisfies the Coulomb condition in 
the interior and remains consistent with the given magnetic field, both up to 
machine precision levels everywhere.

\section{Ghost cell treatment}
\label{A:ghost}

\IGM explicitly creates ghost cells outside the grid.  However, the cell-by-cell 
and global linear algebra methods do not take this into account.  They treat all 
the data as if they are part of the physical grid.  Without any adjustment, this 
causes inconsistencies when data are fed back into \IGM because by default, \IGM 
linearly extrapolates data in the ghost cells.  This is an issue outside of the 
$\mathbf{B}$ to $\mathbf{A}$ calculation; it is purely related to the interface 
with \IGM.  Therefore it is valid to handle the ghost cells independent of the 
$\mathbf{B}$ to $\mathbf{A}$ calculation.

To achieve matching behavior at the boundaries between our methods 
and \IGM, it is necessary to treat different ghost zones in different ways.  
This is due to staggering of the data in our methods and \IGM, as 
described in Section \ref{stagger}.  If we just extrapolated all $\mathbf{A}$-values 
linearly to match the \IGM default, we would introduce errors into the condition 
that $\mathbf{B}=\nabla\times\mathbf{A}$.  Therefore we use hybrid quadratic-cubic 
conditions.  In this scheme, ghost values are handled as follows: normal 
$\mathbf{A}$-field components are extrapolated quadratically from data within the 
physical grid, and tangential $\mathbf{A}$-field components are extrapolated cubically.

For example, if we are looking at the ghost cells on the $+x$-side of the grid, then we would have
\begin{align}
A^x(i,j,k) =& 3 A^x(i-1,j,k) - 3 A^x(i-2,j,k) + A^x(i-3,j,k) \nonumber \\
A^y(i,j,k) =& 4 A^y(i-1,j,k) - 6 A^y(i-2,j,k) + 4 A^y(i-3,j,k) \nonumber \\
           &- A^y(i-4,j,k) \nonumber \\
A^z(i,j,k) =& 4 A^z(i-1,j,k) - 6 A^z(i-2,j,k) + 4 A^z(i-3,j,k) \nonumber \\
           &- A^z(i-4,j,k) \, ,
\end{align}
where, if the number of ghost cells on either side of the grid is $\nghost$, $i$ runs 
from $L+\nghost-1$ to $L+2(\nghost-1)$, inclusive, and $j$ and $k$ cover their full 
range of values, including ghost cells.

\section*{References}

\bibliography{ms}{}

\end{document}